\documentclass[pre,aps,twocolumn,showpacs]{revtex4}
\usepackage{epsfig}
\include{graphics}
\begin{document}

\title{Path-dependent course of epidemic: are two phases of quarantine better than one?}

\author{Varun Nimmagadda}
\affiliation{Novi High School, Novi, MI 48375, USA}
\affiliation{Department of Physics, Oakland University, Rochester, MI 48309, USA}
\author{Oleg Kogan}
\affiliation{Physics Department, California Polytechnic State University, San Luis Obispo, California 93407, USA}
\author{Evgeniy Khain}
\email{khain@oakland.edu}
\affiliation{Department of Physics, Oakland University, Rochester, MI 48309, USA}

\begin{abstract}
The importance of a strict quarantine has been widely debated during the COVID-19 epidemic even from the purely epidemiological point of view. One argument against strict lockdown measures is that once the strict quarantine is lifted, the epidemic comes back, and so the cumulative number of infected individuals during the entire epidemic will stay the same. We consider an SIR model on a network and follow the disease dynamics, modeling the phases of quarantine by changing the node degree distribution. We show that the system reaches different steady states based on the history: the outcome of the epidemic is path-dependent despite the same final node degree distribution. The results indicate that two-phase route to the final node degree distribution (a strict phase followed by a soft phase) are always better than one phase (the same soft one) unless all the individuals have the same number of connections at the end (the same degree); in the latter case, the overall number of infected is indeed history-independent. The modeling also suggests that the optimal procedure of lifting the quarantine consists of releasing nodes in the order of their degree - highest first.
\end{abstract}


\maketitle

\section{Introduction}

Investigating disease dynamics on networks has received much recent attention \cite{Newman book,Another book,Keeling_and_Rohani,Newman PRE,Kenah_and_robbins,Recent_RMP,When_networks,Community_structure,Community_structure2}. Network paradigm has been used in the analysis of epidemics of HIV \cite{Keeling_and_Rohani}, SARS \cite{SARS}, COVID-19 \cite{COVID1}, and flu \cite{FLU}. In many cases in which disease is studied on networks, each node in a network represents a single individual and edges between the nodes represent the contacts between people. The degree of a node is just the number of edges connected to this node; in general, a network has nodes of various degrees, since there are individuals with a high number of contacts and individuals with a low number of contacts. Once the node degree distribution is specified, one can analyze the spread of the disease by employing one of the compartmental epidemiological models, for example, an SIR model \cite{mathbiology}. In this model, the node can be in one of three possible states: susceptible (S), infected (I) and recovered (R). If an infected node is connected to a susceptible node, the latter becomes infected with a certain probability (per unit time). Using an algorithm called a configuration model, it is possible to create a network with any desired degree distribution \cite{Newman book}, \cite{Newman PRE}. Important results for the spread of an epidemic on such networks were obtained in the framework of the SIR model \cite{Newman book}. In particular, there is an exact mapping between the spread of disease on networks and bond percolation.  Therefore, there exists a threshold: the epidemic spreads if the transmission rate exceeds a certain value and decays otherwise. In addition, once the epidemic spreads, one can compute the overall number of people who catch the disease in the course of the epidemic \cite{Newman book}.

This theory assumes that the network itself does not change with time, so that people have the same number of contacts no matter how serious the epidemic situation is.  Recently, a two-way coupling between the network topology and disease dynamics has been studied \cite{Adaptive1}, \cite{Adaptive2}.  However, during the epidemic such as COVID-19, another temporal variation of the network structure has been crucial - one that results from from the various quarantine measures that were introduced (and are currently in effect) in many countries across the globe.  A quarantine aims at reducing the number of contacts, and therefore, the node degree distribution (1) differs from the normal life situation \cite{Quarantine1} and (2) changes once one phase of the quarantine (say, the strict quarantine) is replaced by a second phase of the quarantine (the softer one).  Recently, several authors incorporated quarantine into COVID-19 modeling \cite{Quarantine2,Quarantine3,Quarantine4}.  However, these models are not network-based.  In addition, they do not address the question of the optimal strategy of easing the quarantine in order to minimize the net number of infected individuals - one of the central questions in the present paper.

Looking from a strictly epidemiological point of view, the necessity of a strict quarantine has been called into question \cite{Quarantine_critique1,Quarantine_critique2,Quarantine_critique3}.  One argument against strict lockdown measures is that once the strict quarantine is lifted, the epidemic spread returns to full effect, resulting in the total number of infected individuals during the entire epidemic staying the same. Some argue, therefore, that the way Sweden handled the epidemic \cite{Sweden} might be better. Indeed, if the overall number of infected is the same, softer quarantine measures are more appealing rather than the two phases (a strict quarantine followed by a soft one) from the economic and social considerations. Is this argument correct from a purely epidemiological point of view? We test this idea by running simulations of the SIR epidemic on a network where the node degree distribution changes with time, following administrative decisions to move from a strict quarantine to a soft one. We focus on the final number of susceptible individuals (or the overall number of infected individuals throughout the epidemic) and test different strategies of lifting the quarantine.

\section{Equal number of connections}

Let us first discuss the argument in favor of a single-phase soft quarantine. One can argue that some quarantine is certainly needed to ``flatten the curve". Indeed, without any quarantine measures, the instantaneous number of infected might be so high that it would exceed the medical system capacity. However, one continues, there is no need for a quarantine stricter than that needed to avoid medical collapse. Since eventually the strict quarantine needs to be lifted, the disease will be back, and the total number of people who catch the disease in the course of epidemic remains the same. To support this argument, one can consider the standard SIR model \cite{mathbiology} of an epidemic. The equation for the fraction of infected individuals $I(t)$ reads
\begin{equation}
\frac{dI}{dt} = r\,I\,S - \alpha\,I,
\end{equation}
where $S(t)$ is a fraction of susceptible individuals, the parameter $r$ characterises the transmission rate and the parameter $\alpha$ is the rate of recovery. The number of infected increases if a susceptible person becomes infected (see the first term on the right-hand side) and decreases if an infected person recovers (see the second term on the right-hand side). From this equation, the number of infected starts declining when $dI/dt$ becomes negative, which happens only after the fraction of susceptible individuals becomes smaller than some threshold $S^*$ given by $S^* = \alpha/r$. This means that the epidemic starts decaying only after a certain fraction of population $1-S^*$ (a rather big fraction in the case of COVID-19) catch the disease. Therefore, any quarantine measures can only delay this ``herd immunity", since after lifting the quarantine, we can not avoid infecting less than $1-S^*$ fraction of the population, which often amounts to $70$ or even $80$ percent of the population (depending on the estimates of $r$ and $\alpha$).

Let us now test this argument, considering the SIR model on a network. In general, modeling the spread of the disease on a network is a substantial improvement \cite{Newman book} of the standard continuum (mean field) modeling in two respects. The first advantage is related to the rate of recovery of an infected individual. The continuum modeling implies a Poisson process, which means an exponential distribution of individual disease duration. This is in contrast to observations showing that the distribution is peaked around an average disease duration. Network modeling can easily account for this peaked distribution of individual disease durations. The second drawback of the continuum modeling (that can be eliminated by using networks) is the assumption of an equal number of contacts for each individual, i.e. ignoring the underlying microscopic structure of the social network \cite{Newman book}.

Let us first eliminate only the first drawback and consider a network where once the quarantine is lifted, all of the individuals have an identical number of connections. To be specific, let us consider a somewhat extreme example. Suppose that during the strict quarantine only $10$ percent of the population are ``essential workers" and maintain a high number of contacts, $20$ per day. The remaining population works from home, where each individual has a minimal number of contacts, only $1$ per day. Specifying the transmission probability (taken to be $0.01$) and the duration of the disease (taken to be $14.5$ days), we can follow the number of susceptible as a function of time. To run the simulations, first a network with $10000$ nodes was constructed using the configuration model. Then one node was randomly infected and the SIR model was simulated using the standard Monte-Carlo method. The entire procedure was repeated many times (of the order of $100$ runs) and the average values for $S(t)$ and $I(t)$ were computed taking into account only the runs where an outbreak did occur.

Simulation results are shown in Figure 1. The blue dash-dotted curve shows the fraction of susceptible individuals as a function of time. After several months (once the epidemic is almost over), the quarantine is lifted, so that each node has the same degree $k=20$. The inset shows the fraction of infected as a function of time, and the corresponding blue dash-dotted line describes two waves of the epidemic: the smaller one during the quarantine and the larger one after the quarantine is lifted. Suppose now, we handle the epidemic differently and do not introduce the strict quarantine at all. The node degree distribution is the same as the final node degree distribution above: each node has the same degree $k=20$. The black solid curve shows the fraction of susceptible individuals, while the red dotted line shows the theoretical prediction for the final fraction of susceptible for the SIR epidemic on a configuration model of a network (the mathematics is discussed in the next section). One can see that not only does the epidemic end much earlier, but also the final fraction of susceptible is almost the same as with the strict quarantine! So, the strict quarantine did not change the total number of people who caught the disease in the course of the epidemic. This result is in agreement with the argument based on the continuum SIR model. Does it mean that the strict quarantine is not needed at all? The next section shows that this result is generally wrong and is related to the unrealistic assumption of an equal number of contacts across the population.

\begin{figure}[ht]
\begin{center}
\includegraphics[width=3.4 in]{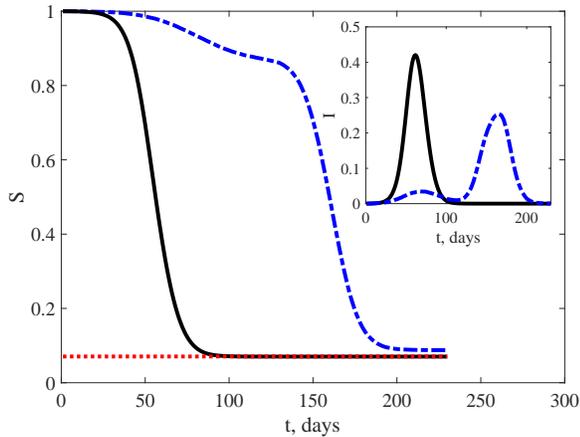}
\caption{The fraction of susceptible individuals as a function of time with and without strict quarantine. Shown are the simulation of the SIR model on a network (the configuration model) with $10000$ nodes. The black solid line corresponds to the no-quarantine case, where all the nodes have the same degree $k=20$. The blue dash-dotted line corresponds to the two-phase scenario. In the first phase (the quarantine), $10$ percent of nodes have degree $k=20$ and $90$ percent of nodes have degree $k=1$. After several months (once the epidemic is almost over), the quarantine is lifted (the second phase), and each node has the same degree $k=20$ as in the no-quarantine scenario. The inset shows the fraction of infected as a function of time for both scenarios. The dotted red line shows a theoretical prediction for the final fraction of susceptible individuals for a single phase (without the strict quarantine), see text.
\label{fig:samedegree}
}
\end{center}
\end{figure}

\section{Distribution of the number of connections}

Now we are going to eliminate the second main drawback of standard continuum (mean-field) models and consider a network where different nodes have a different number of connections. Again we will test the results of two ways of handling the epidemic; importantly, the final node degree distribution is the same in both cases. The first scenario does not imply a strict quarantine: $90$ percent of nodes have degree $14$ and $10$ percent of nodes have degree $50$. The second scenario has two phases, each with its own node degree distribution. The first phase is a strict quarantine: $90$ percent of nodes have degree $2$ and $10$ percent of nodes have degree $50$. These number of contacts are similar to the numbers of casual contacts suggested in the literature \cite{Marissa}. The second phase corresponds to lifting the strict quarantine: $90$ percent of nodes have degree $14$ and $10$ percent of nodes have degree $50$ as in the first scenario. What will be the final fraction of susceptible individuals in these two scenarios?

\begin{figure}[ht]
\begin{center}
\includegraphics[width=3.4 in]{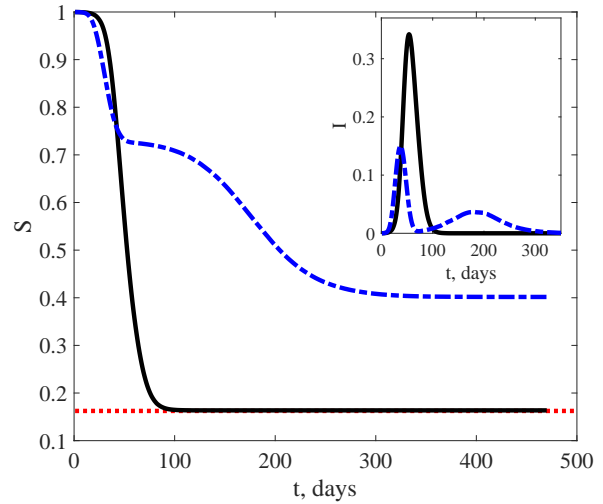}
\caption{The fraction of susceptible individuals as a function of time with and without quarantine. Shown are the simulation of the SIR model on a network (the configuration model) with $10000$ nodes. The black solid line corresponds to the no-quarantine scenario, where $10$ percent of nodes have a high degree ($k=50$) and $90$ percent of nodes have a moderate degree ($k=14$). The blue dash-dotted line corresponds to the two-phase scenario. In the first phase (the strict quarantine), $90$ percent of nodes have a low degree $k=2$ and $10$ percent of nodes have a high degree $k=50$. In the second phase, the quarantine is lifted, and the node degree distribution matches this of the first scenario: $10$ percent of nodes have a high degree ($k=50$) and $90$ percent of nodes have a moderate degree ($k=14$). The inset shows the fraction of infected as a function of time for both cases.
\label{fig:notthesamedegree}
}
\end{center}
\end{figure}

Figure 2 shows that the results are completely different. While the final fraction of susceptible individuals without the strict quarantine is $S_f \approx 0.16$, the final fraction of susceptible in the scenario with quarantine is $S_{fq} \approx 0.4$, more than two times larger. This means that the quarantine helped more than $20$ percent of population, who avoided catching the highly unpleasant disease. This also illustrates a very interesting point: the final steady state depends not only on the final node degree distribution (which is the same in both scenarios), but also strongly depends on the history.

The final fraction of susceptible individuals for a single phase (without quarantine) can be computed theoretically. We present here only a summary of results, and the reader is referred to Ref. \cite{Newman book} for more detail. The calculation uses the concept of a generating function for the node degree distribution defined as $g_0(z) = \sum_{k=0}^{k=\infty} p_k z^k$, where $p_k$ is a fraction of nodes with degree $k$. In our case, the generating function is given by $g_0(z) = 0.9\,z^m + 0.1\,z^{50}$. The generating function for the excess degree distribution is $g_1(z) = (1/\bar{k})dg_0/dz =(0.9\,m\,z^{m-1} + 5\,z^{49})/(0.9\,m+5)$, where $\bar{k}=0.9\,m+5$ is the average degree. Solving numerically the equation $x = 1 - \phi + \phi g_1(x)$, we obtain $x$ and compute the final fraction of susceptible individuals for a single phase, $S_f = g_0(x)$. Here $\phi = 1-exp(-\beta \tau)$, where we considered the following transmission rate and disease duration: $\beta = 0.01$ and $\tau=14.5$ days. The computed value of $S_f$ (for $m=14$, $S_f = 0.1624$) is shown in Figure 2 by the dotted red line. A similar calculation for $S_f$ was performed for the node degree distribution presented in Figure 1 (see the dotted red line). It is possible to perform a theoretical calculation for the fraction of susceptible $S_k$ and infected $I_k$ nodes of a certain degree $k$ as a function of time. Using a degree-based approximation of the SIR model \cite{Newman book}, one can write down rate equations for $S_k$ and $I_k$. These equations however, do not accurately reproduce the results of Monte-Carlo simulations on the network (shown in Figures 1 and 2) since they assume exponential distribution of disease duration times \cite{delay}.

\begin{figure}[ht]
\begin{center}
\includegraphics[width=3.4 in]{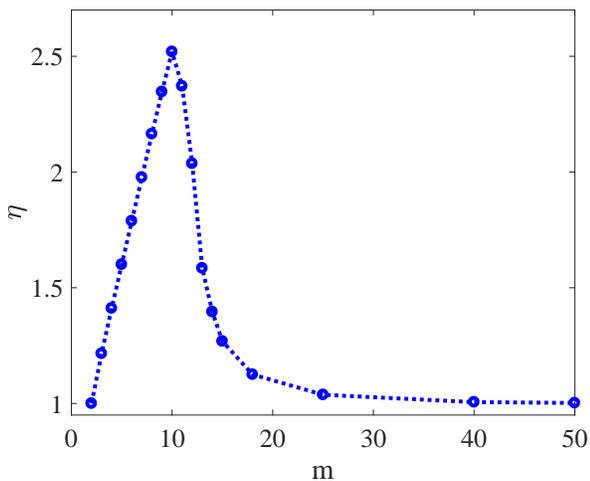}
\caption{The relative importance of the strict quarantine. Shown is $\eta = (1-S_f)/(1-S_{fq})$ as a function of the lower degree $m$, where the final node degree distribution is the following: $90$ percent of nodes have a low degree $m$ and $10$ percent of nodes have a high degree $50$.
\label{fig:effect}
}
\end{center}
\end{figure}

The relative importance of the strict quarantine can be characterised by comparing the final (steady state) fraction of susceptible individuals: $\eta = (1-S_f)/(1-S_{fq})$. Considering the final node degree distribution generated by $g_0(z) = 0.9\,z^m + 0.1\,z^{50}$, we computed $\eta$ as a function of the degree $m$. Figure 3 shows these numerical results. As expected, in the limit of low $m$, $\eta$ tends to $1$, since in this limit, phase $2$ does not differ from phase $1$. As we have shown in the previous section, $\eta$ tends to $1$ also in the limit of high $m$, since the quarantine has no effect for the degree distribution where all individuals have the same number of contacts. The maximum of $\eta$ for intermediate values of $m$ is related to the fact that lifting the quarantine does not lead to an outbreak as most of the high degree nodes caught the disease during phase $1$ and are immune during phase $2$. Therefore, as $m$ increases (from $m=2$ to $m=10$), $S_f$ decreases, but $S_{fq}$ does not change. For larger $m$, the second wave of epidemic becomes stronger leading to a decrease in $S_{fq}$ and smaller values of $\eta$.

\begin{figure}[ht]
\begin{center}
\includegraphics[width=3.4 in]{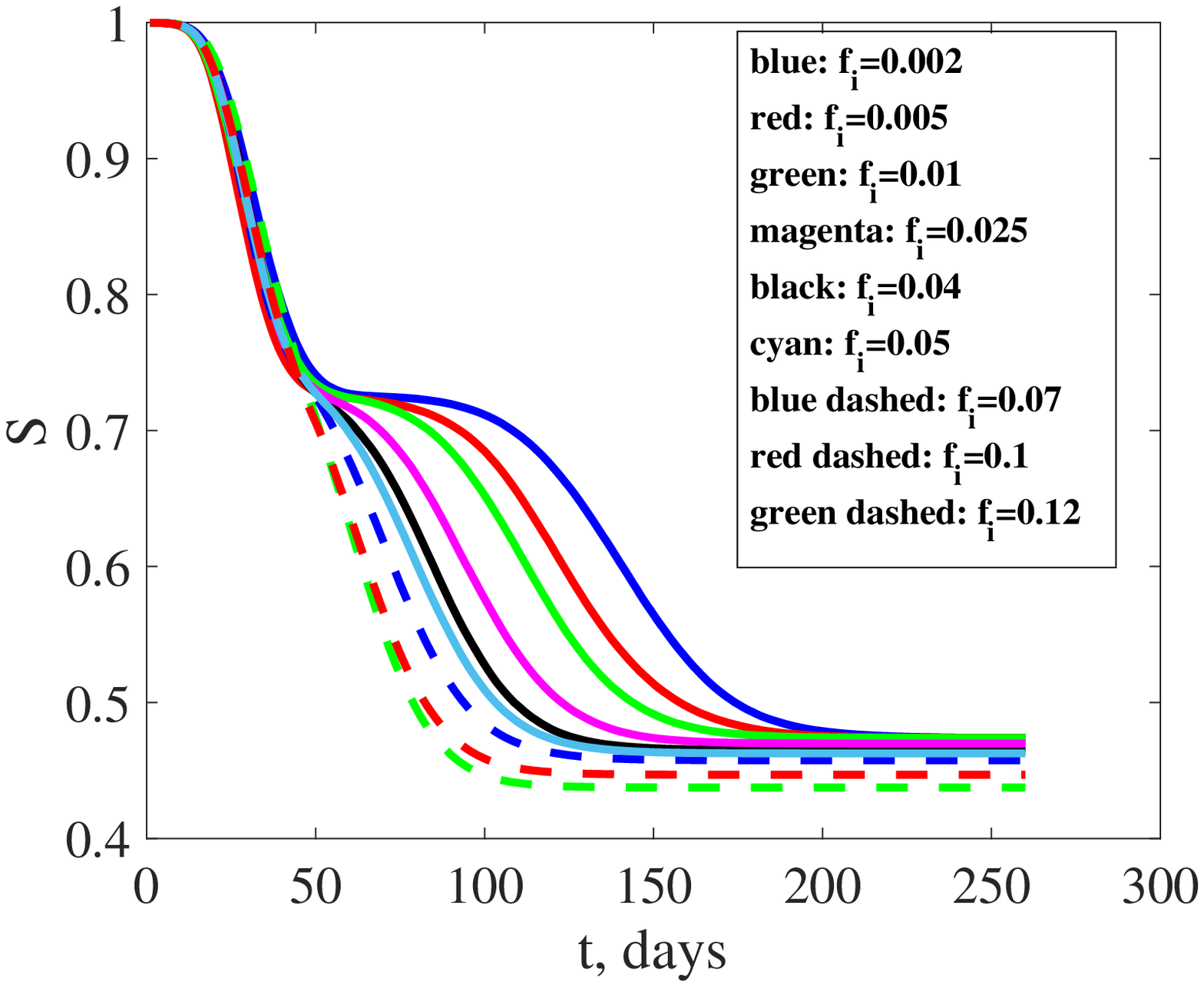}
\includegraphics[width=3.4 in]{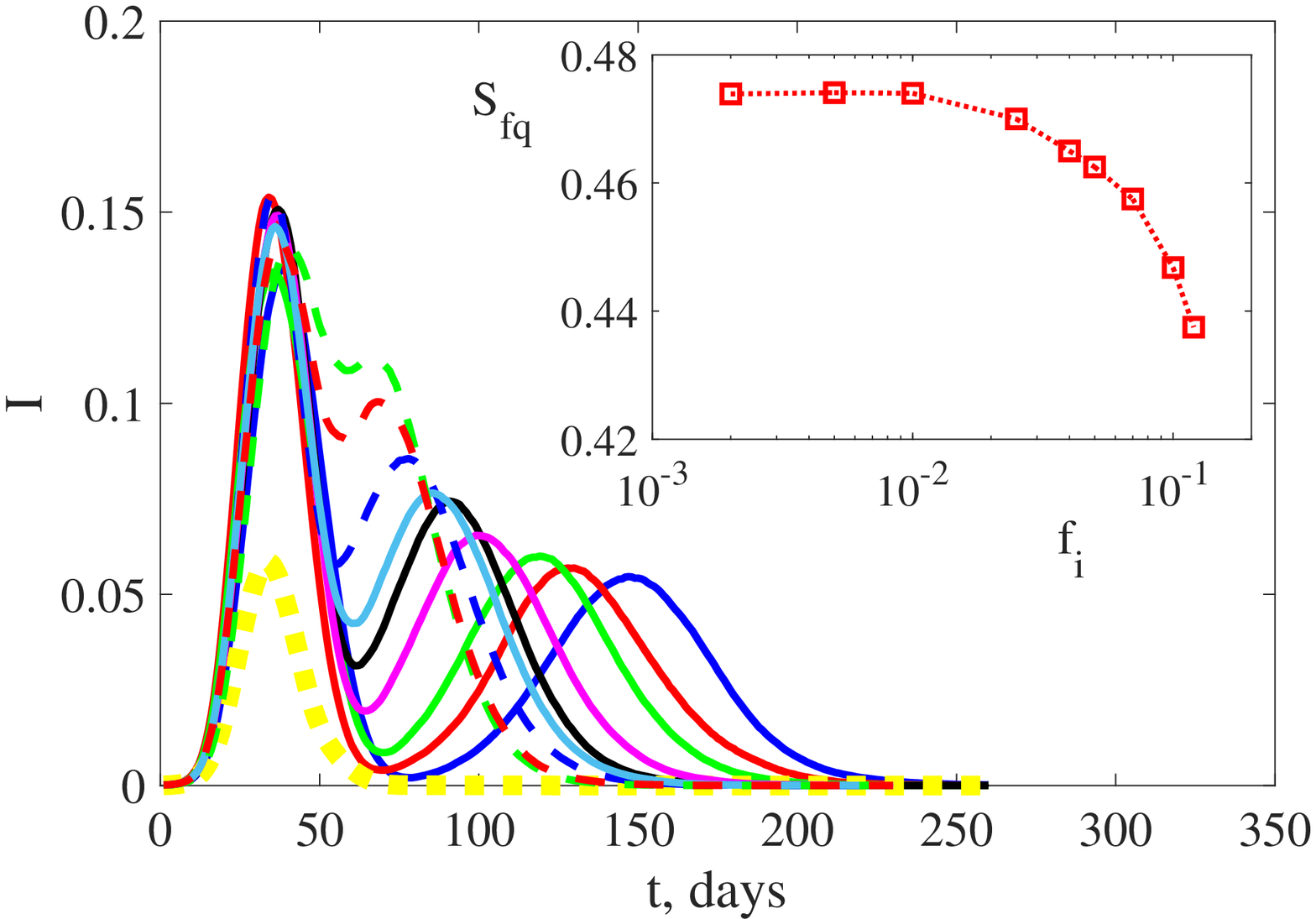}
\caption{The fraction of susceptible individuals (the upper panel) and the fraction of infected individuals (the bottom panel) as a function of time for the two-phase dynamics: the first phase is a strict quarantine, and the second phase is a softer quarantine. See text for the corresponding node degree distributions. The quarantine is lifted when a fraction $f_i$ of the population is still infected. Different curves in both panels correspond to different timing of lifting the quarantine (different values of $f_i$). The inset in the lower panel shows the final fraction of susceptible individuals as a function of $f_i$, demonstrating the effect of timing.
\label{fig:effect}
}
\end{center}
\end{figure}

We have also checked that the effect persists for more complicated node degree distributions. Let us say that during the strict quarantine $90$ percent of nodes have a low degree $2$ and $10$ percent of nodes have a high degree $50$ as before. These high-degree nodes can represent cashiers in big grocery stores \cite{Quarantine1}. Lifting the quarantine means allowing gatherings (for example, $70$ percent of the population increase the number of connections from $2$ to $6$) and opening smaller stores and restaurants etc (for example, $20$ percent of the population increase the number of connections from $2$ to $25$). The numbers are of course quite arbitrary, but they represent the situation qualitatively.  Figure 4 shows results similar to the ones on Figure 2: the quarantine helps. In addition, we investigated the effect of timing: when it is best to lift the quarantine. Reaching the steady state takes time, and it is difficult to keep the population in a strict quarantine for a long time. What happens if we lift the quarantine earlier? Suppose, the quarantine is lifted when a certain (small) fraction of the population (let us denote it by $f_i$) is still infected. The upper panel (as well as the inset in the lower panel) of Figure 4 shows that the second wave of the epidemic is independent of $f_i$ for small values of $f_i$ and only after a while starts growing with $f_i$.

The explanation of this effect can be seen in the lower panel of Figure 4. The strength of the second wave seems to crucially depend on the fraction of the remaining high degree nodes. As depicted by the dotted yellow line in the lower panel of Figure 4, the epidemic in the population of high degree nodes reaches the steady state significantly faster (since these nodes are catching the disease earlier). Once the steady state is reached, the final number of susceptible individuals $S_{fq}$ stops changing (see the inset).

\section{Summary and discussion}
This work discusses the importance of a strict quarantine from the purely epidemiological point of view. One argument in favor of it is clear: it is called  ``flatten the curve" and says that some quarantine measures are needed so that the instantaneous number of infected individuals would not exceed the medical system capacity. However, should we really go beyond that and apply stricter measures?

We have shown that in the case when each individual has the same number of contacts, the strict quarantine is not necessary: the same total number of people catch the disease during the entire duration of epidemic. However, in reality, the node degree distribution is not strongly peaked around a certain degree, and there are individuals with low number of connections (low degree nodes) and individuals with a high number of connections (high degree nodes). In this case, the two phase (or a multi-phase) procedure of the strict quarantine followed by a softer quarantine can save a substantial fraction of population from being infected. The effect results from the fact that essential workers (people who still go to work during the strict quarantine) have a high number of connections (cashiers in the big grocery stores or medical staff are the common examples). A large number of these high degree nodes become infected and then immune by the time the strict quarantine is lifted, decreasing the average degree of the remaining network.

This results suggest an optimal degree-based procedure for lifting the quarantine: high degrees go first. In practice, when the state lifts the strict quarantine (or moves from one phase of quarantine to the next phase), there is always a choice. One can open smaller stores (where cashiers are high-degree nodes) or/and one can allow gatherings (which typically consist of low degree nodes). The model suggests that the stores need to be opened first: this way we can save many individuals (mostly low degree nodes) from being infected.

Clearly, while qualitatively representative of real-life, these results were obtained using a toy model with simplified node degree distributions. To make quantitative predictions, one needs to have data on realistic node degree distributions during different stages of quarantine. Although some attempts to measure the network of contacts in small group experiments are known \cite{nodedegree}, to our knowledge, the real life node degree distributions on a scale of a city are not available. Finally, in this work we did not take into account the spatial degrees of freedom, although the travel patterns are certainly important and are known to be affected by the epidemic \cite{travel}. We have checked however, that the main effect persists in the spatial model of neighborhoods on the lattice \cite{Quarantine1}, and the results obtained with that model (not shown) are similar to the ones presented in Figure 2.

\begin{acknowledgments}

One of us (E.K.) is grateful to Tali Khain for fruitful discussions.

\end{acknowledgments}



\begin{thebibliography}{5}

\bibitem{Newman book} Newman M. E. J. , \emph{Networks, Second Edition} (Oxford University Press, Oxford, 2018).
\bibitem{Another book} Barrat A., Barth\'{e}lemy M., and Vespignani A., \emph{Dynamical Processes on Complex Networks} (Cambridge University Press, Cambridge, 2008).
\bibitem{Keeling_and_Rohani} Keeling M. J. and Rohani P., \emph{Modeling infectious diseases in humans and animals.} (Princeton University Press, 2011).
\bibitem{Newman PRE} Newman M. E. J., \emph{Phys. Rev. E}, \textbf{66} (2002) 016128.
\bibitem{Kenah_and_robbins} Kenah E. and Robins J. M., \emph{Phys. Rev. E}, \textbf{76} (2007) 036113.
\bibitem{Recent_RMP} Pastor-Satorras R., Castellano C., Mieghem P. V., and Vespignani A., \emph{Rev. Mod. Phys.}, \textbf{87} (2015) 925.
\bibitem{When_networks} Bansal S., Grenfell B. T., Meyers L. A., \emph{J. R. Soc. Interface}, \textbf{4} (2007) 879.
\bibitem{Community_structure} Slathe\'{e} M., Jones J. H., \emph{PLoS Comput. Biol.}, \textbf{6} (2010) e1000736.
\bibitem{Community_structure2} Liu Z. and Hu B., \emph{Europhys. Lett.}, {\bf 72} (2005) 315.
\bibitem{SARS} Meyers L. A., Pourbohlol B., Newman M. E. J., Skowronski D. M., Brunham R. C., \emph{J. Theor. Biol.}, \textbf{232} (2005) 71.
\bibitem{COVID1} Block P., Hoffman M., Raabe I. J., Down J. B., Rahal C., Rashyap R., and Mills M. C., \emph{Nat. Hum. Behav.}, \textbf{4} (2020) 588.
\bibitem{FLU} Litvinova M., Liu Q.-H., Kulikov E. S., and Ajelli M., \emph{PNAS}, \textbf{116} (2019) 13174.
\bibitem{mathbiology} Murray J. D., {\em Mathematical Biology}, Springer, New York (2002).
\bibitem{Adaptive1} Gross T., D'Lima C. J. D., and Blasius B., \emph{Phys. Rev. Lett.}, \textbf{96} (2006) 208701.
\bibitem{Adaptive2} Moinet A., Pastor-Satorras R., and Barrat A., \emph{Phys. Rev. E}, \textbf{97} (2018) 012313.
\bibitem{Quarantine1}  Khain E., {\em Phys. Rev. E}, {\bf 102} (2020) 022313.
\bibitem{Quarantine2} GrossB., Zheng Z., Liu S., Chen X., Sela A., Li J., Li D., and Havlin S., \emph{EPL}, \textbf{131} (2020) 58002.
\bibitem{Quarantine3} Trigger S. A. and Czerniawski E. B., \emph{Phys. Scr.}, \textbf{95} (2020) 105001.
\bibitem{Quarantine4} Maier B. F. and Brockmann D., \emph{Science}, \textbf{368} (2020) 742.
\bibitem{Quarantine_critique1} Day T., Park A., Madras N., Gumel A., and Wu J., \emph{Am. J. Epidemiol.}, \textbf{163} (2006) 479.
\bibitem{Quarantine_critique2} Raveendran A.V., Jayadevan R., \emph{Diabetes Metab. Syndr.}, \textbf{14} (2020) 1323.
\bibitem{Quarantine_critique3} The Great Barrington Declaration, https://gbdeclaration.org
\bibitem{Sweden} Vogel G., \emph{`It's been so, so surreal.' Critics of Sweden's lax pandemic policies face fierce backlash}, \emph{Science Magazine}, (6 October, 2020).
\bibitem{Marissa} Renardy M., Eisenberg M., Kirschner D., {\em J. Theor. Biol.},  {\bf 507} (2020) 110461.
\bibitem{delay}  Gomes M. F. C., Goncalves S., {\em Physica A}, {\bf 388} (2009) 3133.
\bibitem{nodedegree} Read J. M., Eames K.T.D.  and Edmunds W. J., \emph{J. R. Soc. Interface}, {\bf 5}  (2008) 1001.
\bibitem{travel} Meloni S., Perra N., Arenas A., Gomez S., Moreno Y. and Vespignani A., {\em Sci. Rep.}, {\bf 1} (2011) 62.

\end{thebibliography}
\end{document}